\documentclass[graybox]{svmult}

\usepackage{mathptmx}       
\usepackage{helvet}         
\usepackage{courier}        
\usepackage{type1cm}        
\usepackage{makeidx}         
\usepackage{graphicx}        
\usepackage{multicol}        
\usepackage[bottom]{footmisc}

\usepackage{natbib}		

\makeindex             

\begin{document}

\title*{Knowledge Graphs and\\Natural-Language Processing}
\titlerunning{Knowledge Graphs and NLP}

\author{Andreas L. Opdahl}
\authorrunning{A.L. Opdahl}

\institute{Andreas L. Opdahl \at 
	University of Bergen, 
	Norway, \email{Andreas.Opdahl@uib.no}}
%

\maketitle

\abstract*{Emergency-relevant data comes in many varieties. It can be high volume and high velocity, and reaction times are critical, calling for efficient and powerful techniques for data analysis and management. Knowledge graphs represent data in a rich, flexible, and uniform way that is well matched with the needs of emergency management. They build on existing standards, resources, techniques, and tools for semantic data and computing. This chapter explains the most important semantic technologies and how they support knowledge graphs. We proceed to discuss their benefits and challenges and give examples of relevant semantic data sources and vocabularies. Natural-language texts --- in particular those collected from social media such as Twitter --- is a type of data source that poses particular analysis challenges. We therefore include an overview of techniques for processing natural-language texts.}

\abstract{Emergency-relevant data comes in many varieties. It can be high volume and high velocity, and reaction times are critical, calling for efficient and powerful techniques for data analysis and management. Knowledge graphs represent data in a rich, flexible, and uniform way that is well matched with the needs of emergency management. They build on existing standards, resources, techniques, and tools for semantic data and computing. This chapter explains the most important semantic technologies and how they support knowledge graphs. We proceed to discuss their benefits and challenges and give examples of relevant semantic data sources and vocabularies. Natural-language texts --- in particular those collected from social media such as Twitter --- is a type of data source that poses particular analysis challenges. We therefore include an overview of techniques for processing natural-language texts.}

\section{What are Knowledge Graphs?}

Knowledge graphs originate from Tim Berners-Lee's vision of a \textit{machine-processable} web of data that would augment the original web of \textit{human-readable} documents~\citep{berners-lee_semantic_2001,shadbolt_semantic_2006}. A central idea is to represent data as graphs, with nodes that represent concrete objects, information, or concepts and with edges that represent semantic relations~\citep{allemang_semantic_2011}.

The most central standard is the Resource Description Framework (RDF\footnote{\url{https://www.w3.org/TR/rdf11-primer/}}), which is the standard way of representing knowledge graphs. An RDF graph consists of \textit{triples}, each expressing that a semantic \textit{resource} (the \textit{subject}) has a particular semantic relation (the \textit{predicate} or \textit{property}) to either a \textit{literal value} or another semantic resource (the \textit{object}). Resources and properties are identified using Internationalized Resource Names (IRN\footnote{Here, we use IRN about Uniform Resource Names (URN) that are extended to the Unicode character set, although it remains more common to use the initialism URN even when Unicode is allowed.}), and literals are typically expressed using XML Schema Definition (XSD) datatypes. A special \texttt{rdf:type} property can be used to state that one resource is the type of another, such as in the triple \texttt{dbpedia:Tim\_Berners-Lee rdf:type foaf:Person} (where we have used standard prefixes \texttt{dbpedia:}, \texttt{rdf:}, and \texttt{foaf:} to shorten the IRNs). Standard formats are available for exchanging RDF files, and the new JSON-LD\footnote{\url{http://json-ld.org}} standard extends JavaScript Object Notation (JSON) with semantic tags so that RDF data as can be easily exchanged through web APIs.

RDF Schema (RDFS\footnote{\url{https://www.w3.org/TR/rdf-schema/}}) extends RDF with terms --- represented as IRNs --- that make knowledge graphs richer and more precise. For example, RDFS defines resource types and properties for expressing that one resource type is a subtype of another (i.e., that toxic fume is a kind of pollution), that one property is a subtype of another (i.e., that being a nurse is a form of being a healthcare worker), and that some property is always used with subjects and objects of specific types (i.e., that only living things can be poisoned). The meaning of RDFS terms is defined through axioms and entailment rules. The Web Ontology Language (OWL\footnote{\url{https://www.w3.org/OWL/}}) offers even more precise semantics and automated reasoning on top of RDFS, but computational complexity grows quickly when datasets become large. Therefore, OWL is most effective for smaller and more specific semantic datasets, called \textit{ontologies}. One important use of ontologies is to precisely define and interrelate the resource types and properties that are used to organise and give meaning to larger knowledge graphs. Such ontologies --- even when they are expressed less formally in RDFS --- are often called \textit{vocabularies} (more about that later).

SPARQL (Simple Protocol and RDF Query Language\footnote{\url{https://www.w3.org/TR/sparql11-overview/}}) lets users and programs extract information from knowledge graphs. The result can be tables of information, yes/no answers, or new knowledge graphs. SPARQL Update also lets users and programs modify knowledge graphs by adding or removing triples. SPARQL is supported both by native RDF database management systems, called \textit{triple stores}, and by wrappers that expose tabular and other data in legacy databases as knowledge graphs --- whether as downloadable RDF files, through online \textit{SPARQL endpoints}, or by other means.

The Linked Open Data (LOD) principles offer further advice for creating and sharing knowledge graphs~\citep{bizer_linked_2009}. The four central principles are: 
\begin{enumerate}
    \item sharing graphs using standard formats and protocols such as RDF, RDFS, OWL, and SPARQL; 
    \item using Internationalized Resource Names (IRNs) to name resources (nodes) and properties (edges); 
    \item making these IRNs into dereferencable Internationalized Resource Identifiers (IRIs\footnote{IRIs are Uniform Resource Identifiers (URIs) that are extended to the Unicode character set. They both \textit{name} a resource uniquely and specify its \textit{location} on the web.}) that can be accessed on the web to provide further information about the resource in RDF format; and 
    \item using standard IRNs that are defined in vocabularies as types and properties in graphs.
\end{enumerate}\noindent
Today, more than 1200 datasets that adhere to these principles are openly available in the LOD cloud~\citep{mccrae_linked_2018}, adding up to almost 150 trillion triples. Much-used datasets we will mention later (such as DBpedia, GeoNames, LinkedGeoData, and Wikidata) act as hubs that tie these linked open datasets even more tightly together by offering standard names (again IRNs) for individual people, organisations, places, works, and so on.

Knowledge graphs can also be stored and processed using property graph databases and other technologies outside the semantic standard but, even for such graphs, RDF and SPARQL are commonly used for information exchange.

\section{Benefits and Challenges}

In an emergency situation, diverse data sources must be recombined and used to support complex querying, processing, and reasoning in unforeseeable ways. This is exactly the type of situation where knowledge graphs shine, because they leverage an interoperable set of semantic technologies and tools for quickly and easily interpreting, combining, analysing, and presenting potentially related datasets from different sources. 

\subsection{Benefits}
Given that the right competencies, tools, and infrastructure are in place, knowledge graphs building on semantic technologies and tools have the potential to simplify and speed up all stages of emergency data processing. 
\textit{Identifying} data sources is made easier by semantic search engines and semantically searchable registries of open data (such as \url{http://lod-cloud.net}). 
\textit{Harvesting} semantic data is made easier by standard data-exchange formats such as Turtle, NT and OWL/XML for downloading files, JSON-LD for web APIs, and SPARQL for database endpoints. 
\textit{Lifting} non-semantic data to RDF format is supported by tools such as Karma\footnote{\url{http://usc-isi-i2.github.io/karma/}}, and JSON data from web APIs can be easily lifted to JSON-LD by adding simple semantic metadata. A wide range of wrappers, such as D2RQ\footnote{\url{http://d2rq.org/}}, provide SPARQL access to relational and other DBMSs that do not natively support SPARQL. 
\textit{Identifying vocabularies} to use for lifting is made easier by semantically searchable registries such as Linked Open Vocabularies (LOV\footnote{\url{https://lov.linkeddata.es/dataset/lov}}~\citep{vandenbussche_linked_2017} and LODstats~\citep{ermilov_linked_2013}). 
\textit{Understanding} data becomes easier for humans when the data attributes are marked up with semantically precise tags from well-defined vocabularies. 
\textit{Alignment} of related terms from different vocabularies (and other kinds of ontologies) is supported by techniques and tools that use term and structural similarity as indicators of term equivalence and of other semantic relations between terms. 
\textit{Recombining} data from different data sets is the most central strength of knowledge graphs: as soon as their vocabularies have been aligned, knowledge graphs can be recombined simply by loading them into the same triple store or through SPARQL, using federated queries that combine partial results from multiple endpoints. 
\textit{Enriching} data means to recombine a dataset with reference data, for example from the Linked Open Data (LOD) cloud. 
\textit{Contextualising} and \textit{validating} data is thus simplified further by openly available semantic datasets that can be used to make data even easier to understand and to control its validity. 
\textit{Reasoning} over data is supported to some extent by the description logic (DL) subset of OWL, although computational effort may grow quickly for large ontologies if they are not carefully designed. Rule-based reasoning is therefore more applicable to large datasets than DL reasoning. 
\textit{Visualising} semantic data, e.g., in dashboards, is also well supported. 
In all these processing stages, the strength of knowledge graphs and semantic technologies lies in the same set of ideas and practices: expressing knowledge uniformly in a standard format (RDF or OWL) that is annotated semantically using well-defined terms (IRIs) defined as part of semantically interlinked vocabularies that are expressed in the same standard formats (RDFS or OWL).

\subsection{Challenges}
A full stack of semantic technologies for knowledge graphs is already available for simplifying and speeding up information processing in an emergency situation. The challenge is to have the right combinations of competencies, capacities, and tools already in place when disaster strikes. 

On the \textit{competence side}, it is critical to recruit and train volunteers with the right combination of semantic-technology competence and collaboration and communication skills. To have maximal impact in an emergency, a semantic technologist must not only be expert in the use of their tools and techniques, but also be able to communicate well with emergency workers and perhaps directly with the people affected. Communicating in an emergency situation is particularly challenging, because the people involved: may be scared, fatigued. and otherwise working in stressful situations; will have a broad variety and levels of other competencies and skills; may come from different cultures, use different languages and perhaps operate in different climates and time zones; may not be knowledgeable and skilled in ICT; may experience low-quality transmission and delays due to long distances and perhaps compromised infrastructures.

On the \textit{capacity side}, most of the semantic interpretation, lifting, combining, and analysing can take place in the cloud in a distributed fashion that makes it highly suitable for volunteer work. Cloud computing platforms such as Amazon's EC2 and others make it possible to set up collaborative computing infrastructures on-demand quickly. The basic tools needed for handling knowledge graphs can be downloaded and installed quickly, and some cloud providers even offer pre-configured virtual hosts (such as Amazon's Machine Images, AMIs) that can be instantiated on demand. Hence, dedicated emergency machine images can be defined in advance where important and trusted reference datasets have already been loaded into a running triple store, along with ready-to-user tools such as as data scrapers and lifters, ontology editors, programming tools and APIs, visualisers, dashboard generators, and various types of social emergency software. Training to create, use, and curate such advance-prepared infrastructures is therefore a useful emergency-preparation activity, and mastering management and use of virtual hosts and other cloud infrastructures is a useful competence.

On the \textit{tool side}, for all types of non-semantic data, precise semantic lifting is essential to avoid information loss. We have already mentioned the computational complexity of OWL reasoning. Indeed, computational complexity is a challenge for graph-based reasoning and pattern matching in general, and it is an important consideration both for native RDF programming and when providing and querying SPARQL endpoints. Although triple-store technologies have been used to store more than a trillion triples in benchmarks, most existing technologies do not scale to the biggest data sizes. An important future challenge is therefore to extend current big-data technologies to also handle semantic data. Finally, knowledge graphs and semantic technologies need to become seamlessly integrated with mainstream machine-learning techniques. 

A final challenge is \textit{textual data}, which must be lifted to semantic form before they can be represented in knowledge graphs. This issue is so central that we will discuss it in a separate section below. 

\section{Vocabularies for Emergency Response}

Semantic technologies, LOD, and knowledge graphs rely heavily on vocabularies, expressed either in RDFS or more precisely and formally as OWL ontologies. Vocabularies define terms that can be used to make the meaning of knowledge graphs explicit, precise, and easier to understand. The terms in a vocabulary provide standard IRNs for the most important resource types and properties in a domain. 
For example, an organisation vocabulary can define resource types for \textit{Person} and \textit{Project} and a \textit{currentProject} property to relate them. We have already mentioned Linked Open Vocabularies (LOV\footnote{\url{https://lov.linkeddata.es/dataset/lov}}), a web site that offers a searchable overview over and entry point into the most used vocabularies. 
Precisely defined and interlinked vocabularies also make it easier to combine knowledge graphs that use different vocabularies. 

There is no all-encompassing and widely accepted ontology that covers all of emergency management. But many data-exchange standards have been proposed for specific concerns, such as people, organisations, resources, infrastructure, processes, disaster description, damage assessment, geography, hydrology, meteorology, and topography. Unfortunately, most standards are defined in plain XML or proprietary formats, and some of them are not even publicly available.

Among the vocabularies that are both open and semantic, MOAC (Management of a Crisis\footnote{\url{http://observedchange.com/moac/ns/}}) combines three types of crisis information used by: (a) traditional humanitarian agencies, (b) disaster affected communities, and (c) volunteer and technical committees for humanitarian data exchange. Accordingly, MOAC is divided into three sections that offer terms (IRNs) for: emergency types, security incidents, and affected populations (emergency management); shelters, water, sanitation, food, health, logistics, and telecommunications (emergency cluster); and who/what/where/when, needs, and responses (who-what-where). Parts of MOAC are supported by the Ushahidi web platform\footnote{\url{https://www.ushahidi.com}} for emergency management.

HXL (Humanitarian eXchange Language\footnote{\url{http://hxlstandard.org/}}) aims to improve information sharing during humanitarian crises without adding extra reporting burdens. It defines hashtags for describing: places, such as geolocations, populated places and administrative units in countries; people and households, such as affected populations, their needs and characteristics; responses and other operations, such as their capacities and operations; crises, incidents and events, including their causes, impacts and severity; and general metadata, such as data provenance, approvals, and timestamps. It offers a broader infrastructure that also comprises training, tools and other materials, including a semantic version of the vocabulary.

EDXL-RESCUER is an attempt to make the XML-based Emergency Data Exchange Language (EDXL\footnote{\url{http://docs.oasis-open.org/emergency/edxl-de/v2.0/edxl-de-v2.0.html}}) standard available as an OWL ontology. EDXL facilitates sharing of emergency information between government agencies and other involved organisations. It offers terms for: alerts, information about events, affected areas, and additional image or audio resources (the common alerting protocol); requesting, responding to, and committing resources (resource messaging); field observations, causality, illness, and management reporting (situation reporting); hospitals, their statuses, bed capacities, facilities, resources, and services (hospital availability exchange); emergency patients (emergency patients tracking); and high-level information modelling (reference information model). 

Other examples of domain ontologies or vocabularies that can be relevant in emergency situations are: km4city (city data), Linked Datex II (traffic), Semantic Sensor Network Ontology (sensors), Ordnance Survey Hydrology Ontology (hydrology), Weather Ontology (meteorology), USGS CEGIS (topography), Ordnance Survey Building and Places Ontology, E-response Building Pathology Ontology, and E-response Building Internal Layout Ontology. These vocabularies can be used alongside general vocabularies for, e.g., time and duration (OWL-Time), locations (geo, GeoNames, LinkedGeoData), people (FOAF, bio), organisations (org, InteLLEO), events (the Event Ontology), provenance (PROV-O), and data rights (CC). 

\section{Semantic datasets for Emergency Management}

The chapter on Big Data has already reviewed many data sources that are relevant for emergency management. Some of them are also available in semantic formats or, at least, have semantic counterparts. 

The LOD Cloud\footnote{\url{http://lod-cloud.net}}~\citep{mccrae_linked_2018} is a searchable portal of more than 1200 interrelated datasets available as knowledge graphs. It contains both general datasets and sets that are specific to emergency-related domains such as geography, government, social networking, and user-generated content. DBpedia~\citep{auer_dbpedia:_2007,bizer_dbpedia-crystallization_2009} is an automated extraction of structured data from Wikipedia (in particular, its fact boxes) into RDF. It describes more than 14 million resources and is available in over a hundred languages. It is one of the most central hubs in the LOD cloud, where it has been standard practice to name people, organisations, works, and so on using their (dereferencable) DBpedia IRIs. Wikidata\footnote{\url{https://www.wikidata.org/wiki/Wikidata:Introduction}} is Wikipedia's sister project for crowdsourcing structured factual information. The idea is that the information in Wikipedia's fact boxes will be extracted from and maintained by the Wikidata project. Hence, whereas DBpedia extracts its data \textit{from} Wikipedia, Wikidata is a supplier of information \textit{to} Wikipedia. It currently contains around 50 million items with unique IRIs, similar to RDF resources. Although Wikidata's knowledge graph is not natively stored and maintained in RDF, the data is available through a SPARQL endpoint and downloadable as RDF files. GeoNames\footnote{\url{http://www.geonames.org/about.html}} is a crowdsourced open repository of more than 10 million geotagged toponyms (geographical names) categorised using a three-level taxonomy with nine letter-coded top-level categories and more than 600 sub-categories. The nine top-level categories are: countries, states, regions... (A); streams, lakes… (H); parks, areas… (L); cities, villages… (P); roads, railways… (R); spots, buildings, farms… (S); mountains, hills, rocks… (T); undersea… (U); and forests, heaths… (V). GeoNames can be browsed online through a map interface. It is also available as RDF and SPARQL and has a web API. It is common in the LOD cloud to name places using their (dereferencable) GeoNames IRIs. LinkedGeoData~\citep{auer_linkedgeodata:_2009,stadler_linkedgeodata:_2012} is an automated extraction of structured data from OpenStreetMap, much as DBpedia is an extraction from Wikipedia. BabelNet\footnote{\url{https://babelnet.org/}} is a multi-lingual word net~\citep{miller1995wordnet}. LODstats\footnote{\url{http://lodstats.aksw.org/}}~\citep{ermilov_linked_2013} has been used to index an even larger body of semantic datasets and endpoints and can be used to search for datasets that use specific RDF types, properties, vocabularies, etc.

The big internet-companies like Google, Facebook, and Amazon also maintain large internal knowledge graphs, although the information is not in general open or always represented using standard semantic formats and protocols. In some cases, commercial data can be sampled or shared in an emergency situation, either pro bono or paid. Google's Emergency Map service and Person Finder\footnote{\url{http://www.google.org/\{crisismap,personfinder\}}}  are examples of such services, although they are not exposed through semantic interfaces.

Google also supports the GDELT project\footnote{https://www.gdeltproject.org/}, which continuously harvests and analyses media in print, broadcast, and web formats in over 100 languages. 
The GDELT Event Database represents and codifies physical events reported in the world news, whereas the GDELT Global Knowledge graph represents the reported people, places, organisations, themes, and emotions.
Both databases are open to the public and incremental updates are available every 15 minutes.
Although the graphs are distributed in tabular form with unique identifiers and well-defined columns, the data are not represented in standard semantic format with IRNs and XSD-typed literals. 
GDELT does not target emergency management specifically, but offers an open-data firehose about human society that can be used to monitor unstable situations and escalating crises.

The new JSON-LD\footnote{\url{http://json-ld.org}} format extends basic JSON in a simple way with semantic tags taken from standard vocabularies. JSON-LD makes it easy to lift JSON-based APIs to a semantic format, so the responses can be inserted directly into knowledge graphs as soon as a suitable vocabulary has been found or created and interlinked. Data represented in XML-based or other formats, such as from Google Person Finder, can easily be converted to JSON before lifting to JSON-LD by adding simple semantic metadata. 

Semantic web APIs also make it much easier to connect the rapidly growing number of more or less smart things available on the internet. Networks of sensors, actuators and other networked devices on the Internet of Things~\citep{atzori_internet_2010} can thereby be identified, integrated, and leveraged much more quickly and easily in an emergency situation, and the information they provide becomes easier to recombine with semantic data from other sources. Smart semantic things can describe, gain access to, and reason about their own context, They can describe themselves and their services semantically in graph form, making them more self-contained and easier to find, for example using the new Semantic Sensor Network Ontology.

Regular datasets that are available as spreadsheets or in SQL databases can also be lifted easily to semantic format.
We have already mentioned Karma\footnote{\url{http://usc-isi-i2.github.io/karma/}}, which is one of several semantic lifting tools that can generate RDF from structured (tabular or hierarchical) data and D2RQ\footnote{\url{http://d2rq.org/}}, which is a much-used wrapper for creating  SPARQL endpoints and RDF interfaces on top of SQL databases. Automatic semantic annotation of images, video, and audio is an emerging area. In particular, deep neural convolution networks have made image analysis much more precise in recent years~\citep{krizhevsky2012imagenet}.

Nevertheless, some of the most important information during an emergency will be available as text, in particular as messages harvested from social media in real time. The next section therefore discusses natural-language processing and lifting of texts into semantic form as knowledge graphs.

\section{Analysing Natural-Language Texts}

\subsection{Pre-processing}
Natural-language processing (NLP) use AI and ML techniques to make the semantic content of written texts processable by computers. Central challenges are to identify: which topics and things a text is about; how the topics and things are related; as well as which attitudes and emotions the text expresses. Conventionally, NLP has built on a pre-processing pipeline that combines all or some of the following steps~\citep[chapter~3]{castillo_big_2016}:
\begin{enumerate}
  \item{\textit{Character decoding and tokenisation} breaks the text into a list of words, word pieces, or even single characters, called \textit{tokens}, that are represented using a standard character set such as Unicode.}
  \item{\textit{Normalisation} standardises use of abbreviations, accents, emoticons, shorthands, slang, upper- versus lower-case characters, etc.}
  \item{\textit{Stopword removal} eliminates words that are too common to convey much meaning, such as ``of'', ``the'', and ``or''. One much-used stopword list contains around 300 words but, for some types of analyses, aggressively eliminating as much as the 20\% most frequent words produce the best results. Removing little used words is also common.}
  \item{\textit{Stemming or lemmatisation} are two alternative ways of handling words such as ``build'', ``builds'', ``built'', ``builder'', and ``building'' that are grammatical forms of the same word (and stem) ``build''. The difference is that stemming uses simple pattern-based string substitutions (typically based on regular expressions), whereas lemmatisation embeds more lexical and grammatical knowledge, including exception lists. For example, a hypothetical and very simple stemmer might treat the word ``was'' as the plural form of (the non-word) ``wa'', whereas a lemmatiser would look up its exception list and identify ``was'' correctly as the past tense of ``is''.}
  \item{\textit{Part of Speech (PoS) tagging} parses sentences to assign words to classes such as nouns, verbs, adjectives, and adverbs. Lemmatisation can sometimes benefit from PoS tags, so the order of steps does not have to be strict. For example, a grammatically-informed lemmatiser would recognise ``building'' as a form of ``build'' when it is used as a verb, but retain the form ``building'' when it is used as a noun.}
  \item{\textit{Dependency parsing} detects how the words and phrases in a sentence are related, for example which noun (phrase) that an adjective modifies, which earlier noun phrase that a pronoun refers to, and which noun phrases that are the subject and object of a verb phrase.}
\end{enumerate}
\noindent
While pre-processing has often relied on hand-crafted algorithms and rules, pre-processing with neural networks and other machine-learning techniques has become more common.

\subsection{Word embeddings}
Natural-language processing techniques are developing rapidly. Google's \textit{word2vec} has trained a neural network to predict which words that occur in which contexts in a 1.6 billion-word corpus~\citep{mikolov_efficient_2013,goldberg_word2vec_2014}. The result is a set of \textit{word vectors}, each of which represents the semantics of a word as a few hundred real numbers. \textit{GloVe} has generated a similar set of word vectors using statistical techniques instead of a neural network~\citep{pennington2014glove}. The vectors generated by word2vec and GloVe can describe word meanings on a very precise level that opens up for new modes of analysis and reasoning. For example, when the vector for the word ``France'' is subtracted from the vector for ``Paris'' and the vector for ``Germany'' is added, the sum turns out to be close to the vector for ``Berlin''. Similar additive relations exist between different grammatical forms of the same stem, so that ``biggest'' -- ``big'' + ``small'' produces a vector similar to the one for ``smallest''~\citep{mikolov_efficient_2013}. But word-vector addition and subtraction does not work equally well for all kinds of relations.

Word-embedding techniques have also been used to generate vectors that approximate the meaning of sentences, paragraphs, and documents~\citep{le2014distributed} and even the nodes (resources) and edges (properties) in knowledge graphs~\citep{ristoski2016rdf2vec}, so that the semantic distance between a word or paragraph and a LOD resource can be approximated by the distance (Euclidian or other) between their vector representations. 
Vector representations of words, sentences, paragraphs, documents, LOD resources, and other semantic phenomena are paving the way for research that may increase the quality of NL processing as word embedding becomes better understood and more widely used. 

Word-embedding approaches often skip all but the first step of the conventional pre-processing pipeline, treating even misspellings and punctuation signs as meaning-bearing tokens. Skipping stemming or normalisation can also improve accuracy because grammatical forms carry semantic information. 

\subsection{Analysis problems}
\textit{Sentiment analysis}, sometimes known as opinion mining, attempts to identify whether a text (or its parts) expresses a positive or negative attitude~\citep{pak_twitter_2016,pang_opinion_2008}. Most sentiment analysers are implemented using supervised machine-learning algorithms. For example, a collection of movie reviews where each text is associated with a numerical ranking can be used to train a regression algorithm~\citep{muller2016introduction}. Emotion analysis uses similar techniques to identify more specific feelings such as joy, anger, disgust, sadness, and fear, both for the text as a whole and for the keywords and phrases it contains.

\textit{Negation analysis} attempts to identify negated parts of a text. Otherwise a sentence like ``I did not find the jokes entertaining.'' could easily be scored as a positive statement: the words ``joke'' and ``entertain'' are both positive, and the rest are neutral or stop words.

\textit{Keyword extraction} attempts to find the most important words and phrases in a text. Conventional keyword analysis uses a bag of words that results from pre-processing steps 1-4. Extraction proceeds by comparing this bag to a large corpus of other pre-processed texts (for example news articles or Wikipedia pages). Good keywords are ones that occur many times in the input text, but are rare elsewhere in the corpus. A suitable measure is term frequency-inverse document frequency (TF-IDF). Word phrases can be extracted in much the same way as keywords, but comparing bags of two- and three-word sequences (called 2- and 3-grams) instead of single words~\citep{sebastiani_machine_2002}. 

\textit{Topic identification} is used to identify topics or themes that are related to a text, but that may not be explicitly mentioned in it. For example, a newspaper article may be related to the Summer Olympic Games although the text does not contain that exact phrase nor a synonym. Machine-learning techniques are much used for this purpose~\citep{muller2016introduction}. 
Latent Dirichlet Allocation (LDA) is a statistical technique that identifies groups of words that tend to occur together in a corpus of texts, under the assumption that each such word group marks a topic or theme that a text can be about.
Word-embedding techniques are increasingly being used to identify and represent the topics of sentences, paragraphs, and documents~\citep{le2014distributed}.

\textit{Classification} is similar to topic identification but, whereas topic identification is open, text classification relies on a closed taxonomy of labels. Standard machine-learning approaches are available for single-label or multi-label classification~\citep{muller2016introduction}, and standard clustering algorithms can be used to establish the initial taxonomy structure. Afterwards, other NL techniques can be used to suggest class labels, although manual curation and labelling is also common.

\textit{Named entity recognition (NER)} attempts to identify the individuals that are mentioned in a text, such as people, companies, organisations, cities, geographic features, etc., usually along with their types. Conventionally, this has been treated as a three-step task. First, the words or phrases that name an individual are identified. Common techniques are gazetteer lists (of known names) and typesetting conventions (such as capital initials) in combination with PoS analysis that identifies nouns. Next, the identified names are disambiguated: does the name ``Bergen'' refer to an American actress, a college football team, or a city in the Netherlands, New Jersey, or Norway? Statistical techniques like LDA can be used here, because each meaning of a name like ``Bergen'' will tend to co-occur with different groups of words. Finally, when the meaning of a name is clear, it is represented in some standard way, preferably linked by an IRN defined in a common Linked Open Data resource. Examples of LOD sets that can be used to define IRNs are the English WordNet (its RDF version), the multi-lingual BabelNet, DBpedia, Wikidata, Geo\-Names, and LinkedGeoData. Keywords and phrases, concepts, and categories/labels can also be semantically linked with IRNs using similar techniques. Recently, neural networks have been applied to all three sub-problems, both separately and in combination.

\textit{Relation extraction} is a challenging area that attempts to identify precise semantic relations between the keywords, phrases, concepts, labels, and named entities that are extracted from a text~\citep{wong_ontology_2012}. For example, when a text mentions a ``hurricane'' near the name of a town, does it mean that the hurricane is approaching, hitting, or passing by? Supervised machine learning has been used to extract specific relations in narrow domains, such as sports results. But general relation extraction using deeper PoS tagging and dependency analysis is an open research area. A new generation of neural-network and word-embedding based joint entity and relation extractors and linkers are producing increasingly accurate (complete and precise) results, often surpassing specialised entity recognisers-linkers and specialised relation extractors-linkers.

\textit{Literal extraction} is a two-step task: first identifying data that constitutes a literal such as a phone number, web address, date or time, and then representing its meaning in a standard way, for example as an IRN or XSD-typed literal string. 

\subsection{Discussion}
With the advent of statistical NL analysers trained on large text corpora, the area of natural-language processing is currently progressing rapidly. But not even advanced machine learning and deep neural networks will be able to handle the more difficult problems of natural-language understanding anytime soon. Such problems include irony, sarcasm, and metaphorical speech that presume a shared pragmatic and social understanding between sender and receiver. Current narrow NL and ML techniques have not yet dealt with these higher levels of communication, which approach the so far unsolved problem of general artificial intelligence. On the other hand, emergencies --- in particular when broken down into particular emergency types (avalanche, derailing, fire, terrorism) --- deal with highly specific domains for which precise NL processors can be trained specifically. Also, during emergencies, people can be expected to use simple and straightforward language that makes NLP easier, with limited use of sarcasm, irony, and metaphor.

In the foreseeable future, general NLP will remain useful but inaccurate. In situations where lives, health, property, and the environment are at stake, we cannot fully trust the results of even the most accurate NL analysers on the single-text level. This applies even more strongly to the kind of short and context-dependent messages people write on social media. Nevertheless, NLP techniques will remain useful in emergency situations in at least two ways:
\begin{itemize}
  \item{They can \textit{provide strategic overviews} by aggregating analysis results over collections of many messages, for example by averaging sentiment and emotion scores and by eliminating concepts and named entities that are not repeated across messages. They can offer answers to questions like: ``In a disaster area, how does the sentiment of tweets that mention food change over time in different locations?'' The hope is that aggregation of many messages will cancel or straighten out single-text analysis errors, but some bias may always remain.}
  \item{They can \textit{suggest potentially actionable insights} by identifying single messages or groups of messages that may contain important tactical information, such as a rapidly approaching fire front, a gas leak, or an entrapment. Semantically categorising a single message as a distress call may not alone justify directing a rescuer or medical worker to a dangerous spot. But it can act as a trigger for further information gathering by automatic or manual means. And it can act as one of several indicators that aid tactical operation leaders in making the best possible decisions based on the available information.}
\end{itemize}

\section{Using a Sentiment Analyser}

A wide range of tools support both sentiment analysis and other NLP techniques. They are available as online services, as downloadable programs, or as APIs that can be used from programming languages such as Python, Java, Scala, and R. Most of them bundle several different analysis techniques together in a single interface. 

We will look more closely at the NLP component of IBM's Watson platform\footnote{IBM Watson offers a free online demo at \url{http://natural-language-understanding-demo.ng.bluemix.net/}, but you must register with IBM Watson to get your own API key.}. Through a web interface, the user enters either a plain text or the URL of a web page. In response, the following features are returned:
\begin{itemize}
  \item{\textit{Keywords and phrases}, ranked by their relevance.}
  \item{\textit{Sentiment} of the text as a whole and for the specific keywords and phrases it contains.}
  \item{\textit{Emotions}, such as joy, anger, disgust, sadness, and fear, both for the text as a whole and for specific keywords and phrases.}
  \item{\textit{Named entities}, such as people, companies, organisations, cities, and geographic features, along with their types, relevance, and occurrence counts.}
  \item{\textit{Concepts} that are related to the text, but that may not be explicitly mentioned in it, ranked by their relevance scores.}
  \item{\textit{Categories} selected from a fixed taxonomy and ranked by their relevance scores: IBM Watson's taxonomy is up to five levels deep with more than a thousand leaf nodes and 23 top categories, such as education, finance, news, science, shopping, and sports.}
  \item{\textit{Semantic roles} that break sentences down into their grammatical and semantic parts.}
\end{itemize}

\noindent
Overall sentiment is scored in the [-1, 1] range, whereas emotions and relevance are [0, 1]-scored. The results are returned in a human-readable web page or as machine-readable JSON. For example, the results of sentiment and emotion analysis may look like this in JSON format:

{\small\begin{verbatim}
{
  "sentiment": {
    "document": {
      "score": 0,
      "label": "neutral"
    }
  },
  "emotion": {
    "document": {
      "emotion": {
        "sadness": 0.029943,
        "joy": 0.056795,
        "fear": 0.025568,
        "disgust": 0.034639,
        "anger": 0.549087
      }
    }
  }
}
\end{verbatim}}

\noindent
Of course, the analyser can be accessed through API calls as well, e.g., from a Python program or from a terminal window using the command-line tool curl: 

{\small\begin{verbatim}
curl -X POST -u "apikey:{your-apikey}"                      \
      "https://{your-api}/analyze?version={your-version}}"  \
     --header "Content-Type: application/json"              \
     --data '{
               "text": "Wildfires rage in Arctic Circle as 
                        Sweden calls for help",
               "features": {
                 "sentiment": {},
                 "concepts": {},
                 "entities": {}
               }
             }'
\end{verbatim}}

\noindent
This command will return JSON results about sentiments, concepts, and entities found in the given newspaper headline. If possible, it will also return a DBpedia IRI for each concept and entity. More specific results can be requested using additional arguments, but a single headline usually contains too little context information to be accurately lifted. 

There is a wide range of similar natural language analysers available, differing mostly in precision and in the range of analyses, metrics, and languages they support. For example, DBpedia Spotlight\footnote{A three-language demo is available at \url{https://www.dbpedia-spotlight.org/demo/}.} returns DBpedia IRIs for topics and named entities found in texts in 12 major languages~\citep{mendes_dbpedia_2011}. The code is open and can be trained and tailored to other languages and more specific domains, such as particular types of emergency situations. The BabelNet\footnote{\url{http://live.babelnet.org/}} analyser returns IRIs for topics and named entities in BabelNet, a multi-lingual version of WordNet. 
NLP services that leverage next-generation NL analysers trained on large text corpora are also appearing. It is likely that the quality of NL analysis tools will continue to improve as word embedding becomes better understood and more neural-network based text-analysis APIs and services become available. 

\section*{Exercises}
\begin{enumerate}
  \item  What is RDF, RDFS, OWL, and SPARQL?
  \item  What is a knowledge graph (RDF graph)?
  \item  Outline the following knowledge graph: \emph{Tim Berners-Lee is a person and an author. He has authored a book with title ``Weaving the Web'', published in 2001. Another person, Mark Fischetti is co-author of this book, which has ISBN 0756752310.}
  \item  What are the benefits of knowledge graphs in an emergency situation?
  \item  And what are the main challenges?
  \item  What is LOD? Give examples of LOD resources that can be useful for emergency management. Where can you go to find more?
  \item  What is a vocabulary in connection with RDFS and OWL? Why are vocabularies important?
  \item  Give examples of vocabularies that can be useful for emergency management. Where can you find more?
  \item  What is TF-IDF? 
  \item  What is LDA?
  \item  What are the main steps in natural-language processing?
  \item  What is a sentiment analyser? Explain its typical outputs.
\end{enumerate}

%
\bibliographystyle{spbasic}              	
\bibliography{References-BDEMTextbook}	

\section*{Acknowledgement}

This is a pre-print of the following chapter: Opdahl, A. L., ``Knowledge Graphs and Natural-Language Processing'', published in Big Data in Emergency Management: Exploitation Techniques for Social and Mobile Data, edited
by Rajendra Akerkar, 2020, Springer International Publishing reproduced with
permission of Springer International Publishing. The final authenticated version is
available online at: \url{http://dx.doi.org/10.1007/978-3-030-48099-8}.

\mbox{}

\noindent
Opdahl, A. L. (2020). Knowledge Graphs and Natural-Language Processing. In Big Data in Emergency Management: Exploitation Techniques for Social and Mobile Data (pp. 75-91). Springer, Cham.

\end{document}